# A Centralized Discovery-Based Method for Integrating Data Distribution Service and Time-Sensitive Networking in In-Vehicle Networks

Feng Luo[1], Yi Ren[1] *, Yanhua Yu[1], Yunpeng Li[1], Zitong Wang[1]

1) School of Automotive Studies, Tongji University, Shanghai 201804, China
*Corresponding author: Yi Ren. E-mail: 2011452@tongji.edu.cn

**ABSTRACT**– As the electronic and electrical architecture (E/EA) of intelligent and connected vehicles (ICVs) evolves, traditional distributed and signal-oriented architectures are being replaced by centralized, service-oriented architectures (SOA). This new generation of E/EA demands in-vehicle networks (IVNs) that offer high bandwidth, real-time, reliability, and service-oriented. data distribution service (DDS) and time-sensitive networking (TSN) are increasingly adopted to address these requirements. However, research on the integrated deployment of DDS and TSN in automotive applications is still in its infancy. This paper presents a DDS over TSN (DoT) communication architecture based on the centralized discovery architecture (CDA). First, a lightweight DDS implementation (FastDDS-lw) is developed for resource-constrained in-vehicle devices. Next, a DDS flow identification algorithm (DFIA) based on the CDA is introduced to identify potential DDS flows during the discovery phase automatically. Finally, the DoT communication architecture is designed, incorporating FastDDS-lw and DFIA. Experimental results show that the DoT architecture significantly reduces end-to-end latency and jitter for critical DDS flows compared to traditional Ethernet. Additionally, DoT provides an automated network configuration method that completes within a few tens of milliseconds.

**KEYWORDS**: Data distribution service, Time-sensitive networking, Quality of service, Service-oriented architecture, Network configuration

## 1 INTRODUCTION

In recent years, automotive technology has continuously evolved and more applications are deployed, such as autonomous driving and vehicle-to-everything (V2X)[1]. Driven by these advancements, large volumes of sensor data, control signals, and V2X messages need to be transmitted, placing higher demands on the bandwidth of IVNs. Therefore, Ethernet has been adopted in IVNs[2]. However, safety-critical applications impose stringent requirements on real-time and reliable data transmission that traditional Ethernet cannot meet.

Since 2021, the IEEE TSN Working Group has continuously released and updated standards that extend traditional Ethernet technology in four aspects: time synchronization, deterministic transmission, reliable communication, and network configuration and management. The works in [3] and [4] confirmed that TSN can meet the communication requirements of ICVs, particularly for infotainment and autonomous driving-related data. Standards related to real-time and reliability, such as IEEE 802.1Qbv and IEEE 802.1CB, are already planned for implementation in IVNs.

On the other hand, with the advancement of ICVs, the value of software in vehicles is increasing, and Software-Defined Vehicles (SDVs) are becoming a new trend[5]. In the future, dynamic changes in user and market demands will require software that can be updated and upgraded throughout the lifecycle of the vehicle. In traditional signal-oriented software development models, the communication matrix was statically defined during the design phase, which severely limited software updates and upgrades. To improve the flexibility and scalability of software deployment, the concept of SOA has been introduced in the automotive field, which decouples software and hardware[6].

Released by OMG in 2004[7], DDS is one of the most commonly used service-oriented communication middleware. In 2018, RTI, a major DDS vendor, announced that AUTOSAR AP R18-03 fully supports DDS network bindings, marking the beginning of DDS adoption in the automotive domain. The research in [8] assessed the feasibility of DDS for automotive applications and found that DDS meets most of the requirements. The work in [9] demonstrated that the use of DDS over Gigabit Ethernet has greater potential compared to FlexRay and traditional Ethernet. To enhance the efficiency of DDS communication, the literature [10] proposed a hardware acceleration method that simplifies software complexity by implementing DDS QoS policies at the hardware level. Additionally, the work in [11] developed an automatic discovery method based on enhanced threshold Bloom filters, which improves the real-time of the DDS discovery phase. Current research confirms the feasibility of DDS in automotive applications, but additional hardware acceleration or software optimization methods are needed to enhance performance.

DDS provides various QoS policies to accommodate communication requirements in different scenarios. However, these QoS policies are insufficient to ensure deterministic communication because DDS cannot control the underlying network. Notably, the QoS requirements for DDS flows are explicitly defined in the QoS policies, which serve as a crucial input for network scheduling in TSN. By combining DDS with TSN, the QoS requirements for DDS flows can be guaranteed through TSN mechanisms. However, there are several challenges in applying DDS and TSN in IVNs:

1) DDS Layer: The extensive features and mechanisms increase the complexity of the DDS software, which implies more resource consumption and uncontrollable software processing latency.

2) TSN Layer: Due to service updates and upgrades, DDS flows are dynamically changing. There is a lack of a method to automate the identification of DDS flows and update the TSN configuration.

To address these challenges, the contributions of this paper are summarized as follows:

1) Lightweight DDS Software: Implements only the QoS policies essential for IVNs, thereby reducing the complexity of DDS software.

2) DDS Flow Identification Algorithm: Automatically identifies changing DDS flows.

3) Communication Architecture DoT: Integrate DDS and TSN and support automated network configuration in IVNs.

The structure of the remaining paper is as follows: section 2 discusses the background and related work on DDS and TSN. Section 3 introduces the design methodology of FastDDS-lw. Section 4 describes the dynamic flow identification algorithm based on a centralized discovery architecture. Section 5 details the operation of the DoT communication architecture, and section 6 evaluates the performance of DoT. Finally, section 7 concludes the paper.

## 2 BACKGROUND AND RELATED WORKS

### 2.1 DDS Overview

The DDS standard defines a Data-Centric Publish-Subscribe (DCPS) model, which describes key DCPS entities in the form of classes[12]. Data is the core of the DDS, and each type of data is defined as a specific Topic. A Domain specifies the boundaries of communication, meaning that data can only be transmitted between entities that belong to the same Domain. The DomainParticipant is a node that participates in DDS communication. Publishing and subscribing of data are implemented by the Publisher and Subscriber created by the DomainParticipant. Publishers and Subscribers create and manage several DataWriters and DataReaders respectively. DataWriters/DataReaders send/receive specific types of data, as they are bound to unique Topics upon creation. To ensure the quality of communication services, the DCPS defines 22 QoS policies that provide mechanisms to ensure real-time and reliable communication, methods for resource management, and so on.

The Real-Time Publish-Subscribe (RTPS) protocol[13] is an underlying transport protocol released by OMG for DDS, which defines the interoperability rules that different DDS implementations must adhere to. RTPS describes a platform-independent model through four modules: Structure, Message, Behavior, and Discovery. This platform-independent model can be mapped to UDP/IP, TCP/IP, or even shared memory to support various underlying transports.

The discovery of DCPS entities is a prerequisite for DDS communication. The DDS standard defines four built-in topics: DCPSParticipant, DCPSPublication, DCPSSubscription, and DCPSTopic, which are used for the discovery of DomainParticipants, DataWriters, DataReaders, and Topics respectively. According to the DDS standard, topic discovery is optional. Therefore, this study does not consider DCPSTopic in the subsequent research.

The Discovery module of the RTPS protocol defines a

Simple Discovery Protocol (SDP), which provides the necessary discovery mechanism for DDS. To ensure interoperability, the RTPS protocol mandates that all DDS implementations must at least support SDP. SDP includes the Simple Participant Discovery Protocol (SPDP) and the Simple Endpoint Discovery Protocol (SEDP). The three built-in topics in the DDS standard are mapped to three different data types in RTPS: SPDPDiscoveredParticipantData(abbreviated as SPDPData), DiscoveredWriterData and DiscoveredReaderData(abbreviated as SEDPData).

Figure 1(a) and 1(b) illustrate the processes of SPDP and SEDP. Initially, each DomainParticipant sends SPDPData to a pre-configured multicast group. Receiving other SPDPData signifies the discovery of a remote DomainParticipant. During the SEDP phase, mutually discovered DomainParticipants exchange SEDPData and perform rule-based matching. Ultimately, only DataWriters and DataReaders with the same Topic and compatible QoS policies can discover each other.

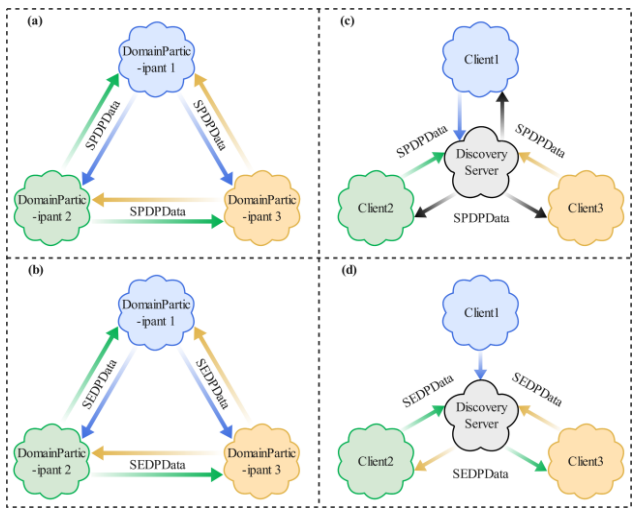

Figure 1. SDP and centralized discovery architecture.

In addition to SDP, some DDS vendors provide extended discovery mechanisms, such as the centralized discovery architecture (CDA) offered by FastDDS[14]. CDA introduces a Discovery Server based on SDP, with all other DDS nodes acting as Clients[15]. Figures 1(c) and (d) depict the workflow of CDA. Unlike SDP, all Client nodes send SPDPData only to the Discovery Server. The Discovery Server then sends SPDPData as a response to the Clients. Upon receiving the response, the Clients continue to send SEDPData to the Discovery Server. The Discovery Server exchanges SEDPData only for mutually matching Clients, assisting them in completing the discovery.

## 2.2 TSN Overview

The TSN standards are designed to ensure real-time and reliable transmission. The Frame Replication and Elimination for Reliability (FRER) mechanism, as defined in IEEE 802.1CB[16], provides reliable transmission services. For reliable transmission, the switch port connected to the Talker enables the replication function, creating a duplicate of the flow. The original flow and its duplicate are then forwarded along two non-overlapping paths. In the switch port connected to the Listener, the elimination function is enabled to identify and remove duplicate flows. FRER provides redundancy in both time and space dimensions[17].

IEEE 802.1Qbv[18] is commonly used to achieve bounded low-latency transmission. The key to the Qbv is time-aware scheduling (TAS). TAS allocates dedicated queues for flows with priorities 0-7 on switch ports. The Gate Control List (GCL) controls the opening times of these queues' gates, ensuring that flows of different priorities are only transmitted during their designated time windows. This global scheduling and planning provide deterministic transmission. TAS requires that the gate opening and closing times across all switch ports strictly adhere to the GCL. Hence, high-precision time synchronization is a crucial prerequisite for TAS.

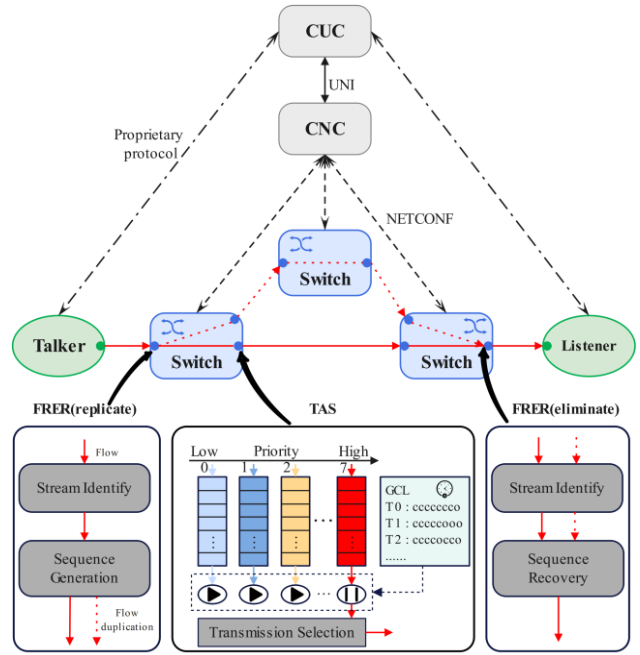

Figure 2. Schematic diagram for IEEE 802.1Qcc, IEEE 802.1Qbv, and IEEE 802.1CB.

FRER and TAS rely on precise and efficient network scheduling, and there has been significant research focused on scheduling issues for in-vehicle TSN. The work in [19] proposed a heuristic scheduling algorithm for in-vehicle TSN, capable of meeting the end-to-end latency requirements of critical streams. The work in [20] introduced four different heuristic scheduling algorithms to address the combined challenges of routing and TAS scheduling. The research in [21] developed a model that integrates TAS with Multi-CQF and used a TSN scheduling optimizer to solve the scheduling issue of mixed traffic in IVNs. To achieve a balance between real-time performance and reliability, four fault-tolerant heuristic algorithms for network scheduling in scenarios where TAS and FRER are jointly deployed[22].

The correct configuration of devices is crucial for implementing various TSN mechanisms. IEEE 802.1Qcc[23] provides methods for network configuration and management. Figure 2 illustrates the fully centralized model defined by the Qcc. The Centralized User Configuration (CUC) and Centralized Network Configuration (CNC) of the control plane are core components of network configuration and management. The TSN Talker/Listener declares the QoS requirements of the flows to the CUC. The CUC delivers flow information to the CNC through the User/Network Interface (UNI). The CNC performs network scheduling and configures the TSN parameters of switches. Once the CNC successfully configures the TSN parameters of the bridges, the TSN mechanisms shown in Figure 2 begin to operate. However, the application of Qcc in IVNs is still in its early stages, and there has been no research specifically targeting the configuration of automotive-grade TSN switches.

**2.3 Related Works**

In the International Organization for Standardization/Open Systems Interconnection Reference Model (ISO/OSI), DDS is a communication middleware between the application and the transport layer, while TSN provides various mechanisms to ensure the QoS at the data link layer. TSN is unaware of the QoS requirements of the DDS, and DDS cannot control the underlying TSN mechanisms. Therefore, an important practical challenge is to integrate DDS and TSN effectively. A part of the current research is dedicated to solving this problem.

The work in [24]designed a Multi-Level TSN based on DDS to transmit synchronized three-phase measurement data, achieving high throughput and low latency. The literature [25] implemented a QoS assurance method in ROS2 by mapping DDS QoS requirements to wireless TSN. The work in [26] tested the end-to-end latency of RTI's Connext DDS. Without TSN, the maximum latency under burst and continuous video stream interference reached 10ms and 5ms, respectively. With TSN enabled, the latency was reduced to under 1 ms. And the research in [27] implemented a fully centralized configuration model. During the system design phase, QoS requirements of DDS flows are written into the CUC, which limits the flexibility of service-oriented communication. In the run-time phase, parameters such as period and VLAN are extracted by parsing DDS flows. Then DDS flows information in JSON format is delivered to the CNC that performs TAS and network configuration. This approach achieved TSN scheduling for DDS traffic. In 2023, OMG released the DDS-TSN protocol[28], defining methods for deploying DDS over TSN. Up to now, no DDS vendors have implemented this protocol.

Current research indicates that deploying DDS over TSN can significantly enhance the QOS of DDS communication. However, related studies are limited, especially in the automotive domain. Furthermore, previous studies have not implemented automated DDS flow identification and network configuration. Therefore, this paper focuses on unresolved issues in the integration of DDS and TSN.

**3 LIGHTWEIGHT DDS IMPLEMENTATION**

Unlike computer networks, most devices in IVNs are resource-constrained. DDS, not originally designed for the automotive industry, often fails to meet the stringent performance requirements of in-vehicle applications. [27] confirmed that the end-to-end latency of OpenDDS is in the millisecond range, which is inadequate for the real-time demands of safety-critical applications. Comparative experiments[29] have shown that FastDDS outperforms OpenDDS and Cyclone DDS in terms of end-to-end latency and throughput. To address these limitations, this study implements FastDDS-lw, a lightweight variant of FastDDS, which reduces complexity and enhances communication performance in resource-constrained IVNs.

**3.1 Analysis of processing latency**

The complexity of DDS software has a direct impact on communication latency. This section quantitatively analyzes the complexity by examining processing latency during both the discovery and publish-subscribe phases.

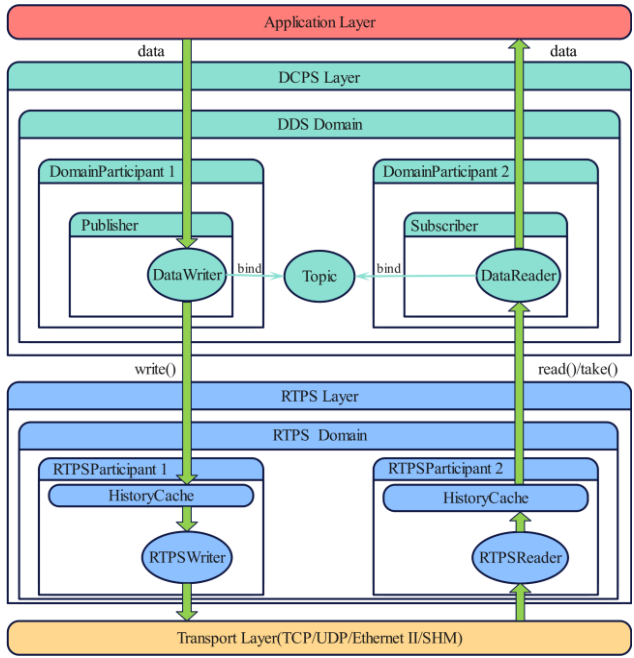

Figure 3. The data publication/subscription process.

Figure 3 illustrates the mapping relationship between DCPS entities and RTPS entities, as well as the data publication/subscription process. On the publishing side, the application first serializes the data and then writes it into the HistoryCache by invoking the $write()$ function. Then all data in the HistoryCache is encapsulated into RTPS messages. Each RTPS message consists of an RTPS Header and several RTPS Sub-messages. Finally, the RTPSWriter transmits RTPS messages to the matched RTPSReader using the associated transport layer protocol. Considering that the TCP/IP retransmission mechanism cannot meet the real-time requirements of IVNs, UDP is adopted in this study. Equation (1) defines the processing latency in the process of data publishing.

$$T_{write} = T_{ser} + T_{hd} + \sum_{i=1}^{m} T_{sub}^{i} + T_{skt\_s} \quad (1)$$

where $T_{write}$ represents the processing latency for data publishing. $T_{ser}$ is the time required for data serialization; $T_{hd} + \sum_{i=1}^{m} T_{sub}^{i}$ is the time required to encapsulate the RTPS message. $T_{skt\_s}$ is the time taken by the socket to send the UDP packet.

Upon receiving a new UDP packet on the socket associated with the DataReader, the RTPS Header along with all RTPS Sub-messages are parsed firstly. Subsequently, the RTPSReader writes the parsed data into the HistoryCache and notifies the DataReader via a callback function. The DataReader then retrieves the data using the $read()/take()$ operation and performs deserialization. Equation (2) outlines the processing latency involved in this data subscription process.

$$T_{read} = T_{skt\_r} + T_{phd} + \sum_{j=1}^{m} T_{psub}^{j} + T_{cb} + T_{dser} \quad (2)$$

where $T_{read}$ is the processing latency during the data subscription process; $T_{skt\_r}$ is the time taken for the socket to receive the UDP packet; $T_{phd} + \sum_{j=1}^{m} T_{psub}^{j}$ is the time required to parse the RTPS message; $T_{cb}$ is the time taken to execute the callback function; $T_{dser}$ is the time required for data deserialization.

DDS software generally establishes a complex multi-threaded architecture. For instance, in addition to the main and receiving threads, FastDDS creates an "Event" thread for each DomainParticipant to manage periodic and time-triggered events. The synchronization mechanisms for thread safety increase the waiting time during thread execution. Consequently, the total processing latency of DDS software $T_{total}$ can be described by Equation (3).

$$T_{total} = T_{disc} + T_{write} + T_{read} + \sum_{k=1}^{n} T_{wait}^{k} \quad (3)$$

where $T_{disc}$ is the processing latency during the discovery phase. $\sum_{k=1}^{n} T_{wait}^{k}$ is the waiting time introduced by the synchronization mechanisms among $n$ threads.

### 3.2 Lightweight DDS Software

For interoperability, FastDDS-lw retains the discovery mechanisms provided by FastDDS. Socket communication utilizes the Asio library, while data serialization and deserialization rely on the FastCDR library. Therefore, the parameters $T_{disc}$, $T_{skt\_s}$, $T_{skt\_r}$, $T_{ser}$ and $T_{dser}$ in Equations (1) to (3) are not the focus of optimization. Apart from these essential communication processes, QoS policies significantly influence the complexity of DDS software. Consequently, this study primarily targets the optimization of QoS policies. Considering the specific requirements of automotive applications and the integration of TSN mechanisms, the following QoS policies have been optimized.

#### 3.2.1 Resource-related QoS policies

The DURABILITY QoS policy manages the lifecycle of data samples and offers four options: 1) data samples are destroyed after being sent to known subscribers; 2) the lifecycle of data samples is the same as the related DataWriter; 3) the lifecycle of data samples aligns with the application; 4) data samples are stored in non-volatile memory. In IVNs, most data require real-time updates, which can be satisfied by option 1). For scenarios requiring historical data retrieval, option 2) allows for DataWriter-level data rollback. Given that options 1) and 2) adequately address the requirements of IVNs, options 3) and 4) are not implemented.

The HISTORY QoS policy specifies the number of data samples that DataWriters and DataReaders can retain. It provides two strategies: 1) retaining all historical data samples, and 2) retaining a specified number of data samples, determined by the "depth" attribute. Considering the limited memory resources in in-vehicle devices, the first strategy is not suitable. With strategy 2), the "depth" attribute can be configured to any value according to the actual requirements of automotive applications.

### 3.2.2 Reliability-related QoS policy

RELIABILITY QoS policy determines the approach to data transmission and offers two modes: "Reliable" and "BestEffort". In the "Reliable" mode, the DataWriter declares data validity through HeartBeat sub-messages, while the DataReader acknowledges receipt of data via AckNack sub-messages. The DataWriter retransmits any valid data not received by the DataReader. Additional sub-messages increase the values of $\sum_{i=1}^{m} T_{sub}^{i}$, $\sum_{j=1}^{m} T_{psub}^{j}$. The time from message loss detection to successful retransmission and receipt may reach the millisecond level. Consequently, the "Reliable" mode is not suitable for IVNs.

Considering the application of the TSN mechanism, the FRER can offer reliable communication. Our experimental results show that, with FRER enabled, the latency for frame replication and elimination in switch SJA1110 is on the order of nanoseconds. Consequently, substituting the "Reliable" mode with the FRER can avoid complex operations and unpredictable communication latency, while still ensuring communication reliability.

### 3.2.3 Real-time related QoS policies

The "period" attribute of the DEADLINE QoS policy specifies the deadline for data publication and subscription. Timed tasks registered with the "Event" thread are created to verify that data is published or subscribed within the deadline. The "duration" attribute of the LATENCY_BUDGET QoS policy defines the maximum acceptable latency for data transmission, which is also monitored by timed tasks. The timed tasks only issue warnings to the application when the "period" or "duration" is reached. However, an increase in the number of timed tasks leads to a larger value of $\sum_{k=1}^{n} T_{wait}^{k}$ in Equation (3).

If the TAS mechanism in TSN is used, it can ensure a deterministic period and end-to-end latency for DDS flows by scheduling all traffic in the network. Therefore, substituting the DEADLINE and LATENCY_BUDGET QoS policies with TAS reduces DDS software complexity while better ensuring real-time DDS communication.

Based on the analysis, QoS policies deemed unsuitable or replaceable by TSN mechanisms are not implemented in FastDDS-lw. However, all attributes of these QoS policies are retained to ensure interoperability. Besides, functions from FastDDS that are not suitable for IVNs, such as XML-based configuration and traffic statistics, have been removed. The shared library file size of FastDDSv2.12.0 is 9.2 MB, while that of FastDDS-lw is 7.9 MB. A detailed comparison of other performance metrics will be presented in section 6.

Table 1. Summary of notations

| Notation | Definition |
|---|---|
| **DDS** | |
| $P$ | Type of RTPS message, $P \in \{pdp, edp, data\}$ |
| $M_P$ | An RTPS message with the type $P$ |
| $T$ | The set of DDS topics in the network |
| $tp$ | DDS topic, $tp \in T$ |
| $W$ | The set of DataWriters in the network |
| $W_{tp}$ | The set of DataWriters with the topic $tp$ |
| $w_i^{tp}$ | A DataWriter with the topic $tp$, $i \in \{1, \dots, |W_{tp}|\}$ |
| $R$ | The set of DataReaders in the network |
| $R_{tp}$ | The set of DataReaders with the topic $tp$ |
| $r_n^{tp}$ | A DataReader with the topic $tp, n \in \{1, \dots, |R_{tp}|\}$ |
| $edp$ | An Endpoint, $edp \in W \cup R$ |
| $qos$ | The set of DDS QoS policies |
| **DDS Flow** | |
| $F$ | The set of flows in the network |
| $F_{tp}$ | The set of flows with the topic $tp$ |
| $f_k^{tp}$ | A flow at the topic $tp$, $k \in \{1, \dots, |F_{tp}|\}$ |

## 4 DDS FLOW IDENTIFICATION ALGORITHM

In the SOA for ICVs, all functions are encapsulated into different services. Flexibility in service deployment is one of the advantages of SOA. Services can be updated and upgraded via OTA technology throughout the vehicle's lifecycle. As a result, traffic within the IVNs is no longer static. Considering that network scheduling in TSN relies on complete and accurate flow information, there is a need for a method that can automatically detect changes in flows and provide the necessary input for network scheduling. In this section, we first briefly describe the model of DDS flows. Then a DDS flow identification algorithm is proposed based on CDA to automatically identify and update the information of DDS flows, including the addition of new flows and the removal of invalid ones. Table 1 lists all the notations used in this section.

### 4.1 DDS Flow Model

Nodes in IVNs deploy various applications to perform specific data transmission tasks. Each transmission task represents a set of RTPS messages carrying DDS Topic information, which constitutes a DDS flow. Each task implements the sending and receiving of DDS flows by creating DDS endpoints (DataWriters and DataReaders). The QoS requirements for each task are specified by the QoS policies associated with the DataWriter, DataReader, and Topic. We assume that designers have defined the transmission tasks required for each node and defined the QoS requirement for each task. In addition, unique address information is assigned to each DDS endpoint. endpoints can be described by Equation (4).

$$edp = \{type, status, guid, topic, locator, qos\} \quad (4)$$

where $type$ represents the type of the endpoint, which can be either "writer" or "reader". $status$ indicates whether the endpoint is active, and $status$ is declared in SEDP messages. $guid$ is the unique identifier of the endpoint in the network. In the RTPS protocol, each entity is assigned a 16-byte identifier known as GUID. $topic$ is the topic bound to the endpoint. $locator$ is the address information of the endpoint, including IP address and port number, thus $locator = \{ip, port\}$. $qos$ specifies the QoS required by the endpoint. This article focuses primarily on the real-time and reliability of the communication, thus $qos = \{partition, priority, deadline, latency, reliability, jitter, size\}$, corresponding to PARTITION, TRANSPORT_PRIORITY, DEADLINE, LATENCY_BUDGET, RELIABILITY, and TSN_EXTENSION QoS policies. PARTITION is used to divide logical partitions within a Domain. TRANSPORT_PRIORITY specifies the priority of the transmission.

It is important to note that no QoS policy is used in the DDS standard to characterize the jitter of the communication delay as well as the byte size of the topic. Therefore, the TSN_Extension QoS policies, as illustrated in Figure 4, are defined in FastDDS-lw. $jitter$ represents the maximum acceptable jitter, while $size$ denotes the total byte size of the topic.

```
1 struct TSNExtentionQosPolicy {
2          Duration_t jitter;
3          unsigned long size;
4 };
5
6 struct Duration_t {
7          long seconds;
8          unsigned long nanoseconds;
9 };
```

Figure 4. TSN_Extention QoS policy in FastDDS-lw.

Each DDS flow $f_k^{tp}$ is described by Equation (5). $tp$ is the topic associated with the flow. $id$ is the unique identifier of the DDS flow. $src$ and $dst$ are the address information of the DataWriter and DataReader respectively. $size$ is the byte size, $vid$ is the VLAN ID, $prio$ is the flow priority, and $prd$ is the flow period. $latency$ denotes the maximum acceptable end-to-end delay, while $jitter$ denotes the maximum acceptable end-to-end delay jitter. $reliability$ indicates the reliability requirements for the flow during transmission. $route$ is the transmission path of the flow. Routing algorithms are not the focus of this paper, so the paths of DDS flows are assumed to be known.

$$f_k^{tp} = \{id, src, dst, size, vid, prio, prd, latency, jitter, \\ reliability, route\} \quad (5)$$

### 4.2 DFIA Based on CDA

The E/EA of automotives is transferring from distributed to centralized[30]. Figure 5 illustrates the zone-oriented centralized E/EA of ICVs. In the distributed architecture, each Electronic Control Unit (ECU) realizes a specific function. To reduce the number of ECUs, in the zone-oriented centralized architecture, the functions are deployed centrally in Zone Control Units (ZCUs). Meanwhile, Central Computing Units (CCUs) with abundant computing resources are used to handle complex tasks such as autonomous driving.

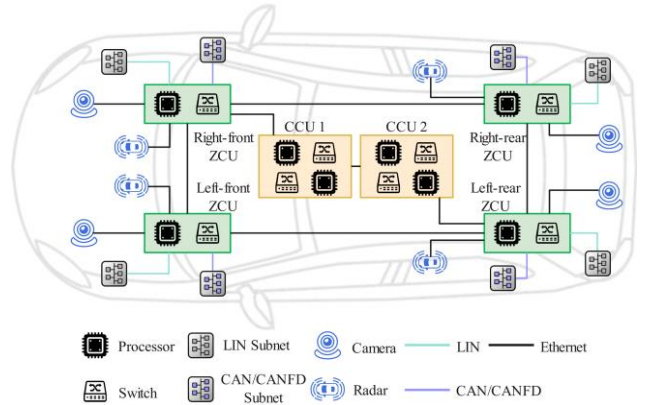

Figure 5. Zone-oriented centralized E/EA.

If CDA is used in the zone-oriented centralized E/EA, the Discovery Server can be deployed in CCUs, and ZCUs act as Clients. CDA offers the following advantages: 1) the matching process of endpoints is handled by CCUs, which can reduce the resource consumption of Clients; 2) compared with SDP, CDA can reduce the network traffic in the discovery phase and save bandwidth for the in-vehicle network; 3) Unlike SDP, which relies on multicast, all the communication between the Client and the Server in CDA is done using unicast, which avoids the security issues introduced by multicast communication; 4) a backup Discovery Server can be set up in IVNs to solve the single point of failure.

In this section, a CDA-based DDS flow identification algorithm is proposed. As the center node of CDA, the Discovery Server can obtain all the SEDP Messages (RTPS messages carrying SEDPData) in the network. Therefore, DFIA is deployed in the Discovery Server to automatically identify changing DDS streams by parsing SEDP Messages and performing rule-based matching and mapping. A learning-capable edge switch has been proposed to monitor flows and extract specific parameters[31]. However, this approach does not ensure QoS during the initial communication phase. DFIA runs in the DDS discovery phase, allowing for the early identification of all potential flows.

```
Algorithm 1: DDS Flow Identification Algorithm
Input: EDP Message M_edp, Flow set F, Writer set W, Reader set R
Output: Flow set F, Writer set W, Reader set R
1  edp ← get_edpinfo(M_edp)
2  tp ← edp.tp
3  if edp.status == UNALIVE then
4      if edp.type == writer ∧ edp ∈ W_tp then
5          W_tp.delete(edp)
6          foreach f_k^tp ∈ F_tp do
7              if f_k^tp.writer == edp then
8                  F_tp.delete(f_k^tp)
9      else if edp.type == reader ∧ edp ∈ R_tp then
10         R_tp.delete(edp)
11         foreach f_k^tp ∈ F_tp do
12             if f_k^tp.reader == edp then
13                 F_tp.delete(f_k^tp)
14 else
15     if edp.type == writer ∧ edp ∉ W_tp then
16         W_tp.append(edp)
17         foreach r_b^tp ∈ R_tp do
18             if check_qos(edp.qos, r_i^tp.qos) then
19                 f_k^tp ← gen_flow(edp, r_b^tp)
20                 F_tp.append(f_k^tp)
21     else if edp.type == reader ∧ edp ∉ R_tp then
22         R_tp.append(edp)
23         foreach w_a^tp ∈ W_tp do
24             if check_qos(w_j^tp.qos, edp.qos) then
25                 f_k^tp ← gen_flow(w_a^tp, edp)
26                 F_tp.append(f_k^tp)
```

Algorithm 1 shows the process of DFIA. The algorithm takes as input SEDP Message $M_{edp}$, the set of flows $F$, the set of DataWriters $W$ and the set of DataReaders $R$. Upon receiving $M_{edp}$, the function $get\_edpinfo()$ extracts the endpoint information $edp$ carried in $M_{edp}$. If $edp.status$ is UNALIVE, it indicates that the endpoint associated with $M_{edp}$ is no longer active. Then the algorithm traverses all the flows in $F_{tp}$ and filters out the ones associated with this endpoint. These flows are considered to be invalid and are subsequently deleted (lines 8 and 13).

Table 2. Mapping rules for endpoint information to DDS flow.

| $edp$ | $f_k^{tp}$ | Mapping Rules |
|---|---|---|
| guid | id | if type == w, id.w = guid |
|  |  | if type == r, id.r = guid |
| locator | src | if type == w, src = locator |
|  | dst | if type == r, dst = locator |
| topic | size | size = fixed_size + qos.size |
| qos | vid | vid = qos.partition |
|  | prio | prio = qos.priority |
|  | prd | prd = qos.deadline |
|  | latency | latency = qos.latency |
|  | jitter | jitter = qos.jitter |
|  | reliability | reliability = qos.reliability |

(Note: fixed_size refers to the total length of static fields, including the RTPS and UDP headers. The DataReader's QoS policy specifies the subscriber's QoS requirements, while the DataWriter's QoS policy defines the publisher's QoS capabilities. Consequently, the DataReader's QoS policy is considered the QoS requirement for the DDS flow.)

An $edp$ is considered a newly added endpoint in the network if $edp.status$ is ALIVE and its associated endpoint does not exist in $W$ and $R$. The $edp$ is first added to the appropriate $W_{tp}$ or $R_{tp}$ based on the topic (lines 16 and 22). Then the rule-based matching algorithm filters out the endpoints that have the same topic and compatible QoS policies (lines 17-18 and 23-24). If matching endpoints are found, the $gen\_flow()$ function generates DDS flows according to the specified mapping rules (lines 19 and 25). Table 2 gives the mapping rules from endpoint information to DDS flow information. The new DDS flows are added to the corresponding subset $F_{tp}$ (lines 20 and 26) to update the flows set $F$. Finally, DFIA outputs the updated set $F$, $W$, $R$.

## 5 DDS over TSN Architecture for IVNs

The lightweight FastDDS-lw proposed in section 3 outperforms in resource-constrained vehicle devices by reducing software complexity. To further enhance the QoS of DDS communication, TSN mechanisms such as TAS and FRER need to be applied. However, DDS and TSN are located in different network layers and cannot perceive each other, which is the biggest challenge for their integrated application. DFIA can automatically identify global DDS flow information, providing a means for DDS and TSN to perceive and interact with each other. Therefore, this section implements a DDS over TSN communication architecture based on FastDDS-lw and DFIA under the zone-oriented centralized E/EA of ICVs.

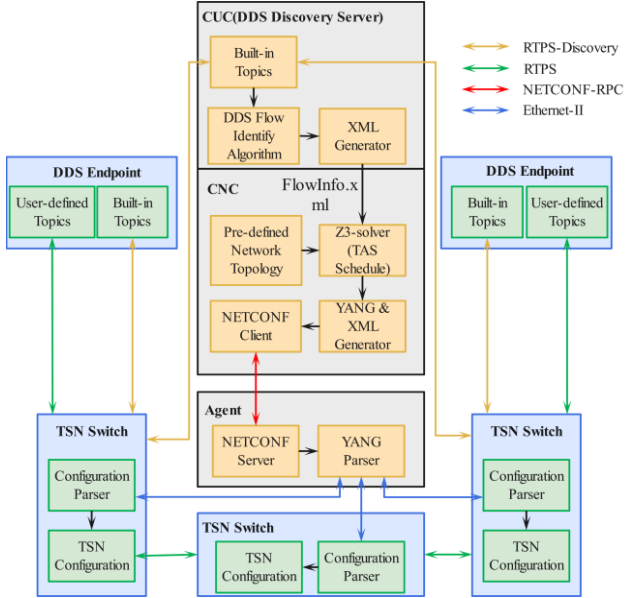

Figure 6. DoT communication architecture.

Since the current in-vehicle switches do not support network management protocols such as NETCONF, the DoT architecture as shown in Figure 6 was constructed by adding a configuration agent node based on the fully centralized model proposed in 802.1Qcc. The operation of DoT is divided into two phases: network scheduling and network configuring, which realize the automated configuration and management of IVNs.

## 5.1 Network Scheduling

Prior to vehicle startup, network scheduling and configuration must be completed to ensure proper communication. Network scheduling requires network topology and information on all flows as input. In the DoT architecture, the DDS Discovery Server integrated with DFIA is deployed in the CUC to identify DDS flows. Information on DDS flows is stored in the FlowInfo.xml file and then delivered to the CNC via the UNI. Figure 7 shows the recorded DDS stream information. At the same time, the CUC notifies the CNC to perform network scheduling through the UNI. Considering the relatively small scale of IVNs and the fact that network topology is likely to change only in case of severe faults, such as link damage or node failures, this paper assumes that network topology is statically defined during the design phase and does not account for changes in the run-time phase.

The CNC is primarily responsible for scheduling all DDS flows through the TAS. The scheduling module of the CNC uses a Satisfiability Modulo Theories (SMT) solver, z3-solver[32], developed by Microsoft Research. DoT is compatible with any scheduling algorithm or solver. The z3-solver uses information about DDS flows and network topology as input and establishes network constraints based on the QoS requirements of DDS flows. It ultimately determines the GCL for switch ports and the transmission time offsets for each stream across different links, ensuring deterministic end-to-end latency for DDS stream transmission. To achieve reliable transmission, the CNC also generates configuration information related to the FRER, i.e., enables flow identification, replication, or elimination functions on specific switch ports.

```
1  <?xml version="1.0" encoding="UTF-8"?>
2  <Flow Features>
3    <Flow1 Flow ID="1">
4      <Topic Name>ExampleTopic</Topic Name>
5      <Size>1454</Size>
6      <Node Info>
7        <Talker Guid>01.0f.c5.1c.31.2c.e4.d7.00.00.00.00|0.0.1.3</Talker Guid>
8        <Talker Address>UDPv4:[192.168.137.181]:7411</Talker Address>
9        <Listener Guid>01.0f.eb.08.63.26.b3.2f.00.00.00.00|0.0.1.4</Listener Guid>
10       <Listener Address>UDPv4:[192.168.137.20]:7411</Listener Address>
11     </Node Info>
12     <QoS Info>
13       <VID>2</VID>
14       <Priority>7</Priority>
15       <Deadline>500000</Deadline>
16       <Latency>500000</Latency>
17       <Jitter>25000</Jitter>
18       <Reliability>Reliable</Reliability>
19     </QoS Info>
20   </Flow1>
21 </Flow Features>
```

Figure 7. Information about DDS flows identified by DFIA.

## 5.2 Network Configuring

All switches and DDS endpoints in the network receive the results of network scheduling and complete the corresponding configuration so that TAS and FRER take effect. To facilitate the transmission, the results of the z3-solver are stored using the YANG data model, and an XML-format TSN configuration file is generated. Then CNC distributes the TSN configuration files to all switch nodes via NETCONF. However, current in-vehicle switches (e.g., the NXP SJA1110 used in this paper) do not support the NETCONF protocol.

To address this issue, an intermediary agent node was designed. A TSN configuration protocol specifically designed for the SJA110, developed by our team, is integrated into both the agent and the switches. First, the agent and CNC interact via NETCONF-RPC. The CNC, acting as the NETCONF Client, transfers the TSN configuration files to the agent's NETCONF Server module. The "YANG Parser" module of agent then parses the TSN configuration files and sends Ethernet II configuration frames to switches. These configuration frames use 0xF123 as the "Ethernet Type" and fill the Ethernet data segment with TSN configuration information. Figure 8 shows the format of the configuration frames specified by the TSN configuration protocol. "TSN Type" indicates the pre-configured TSN protocol. Each switch is assigned a unique "Device ID". "TSN Config" contains specific TSN configuration details, such as GCL related to the TAS. Switches parse the received configuration frames and update the corresponding TSN configuration. Finally switches return the configuration status to the agent node.

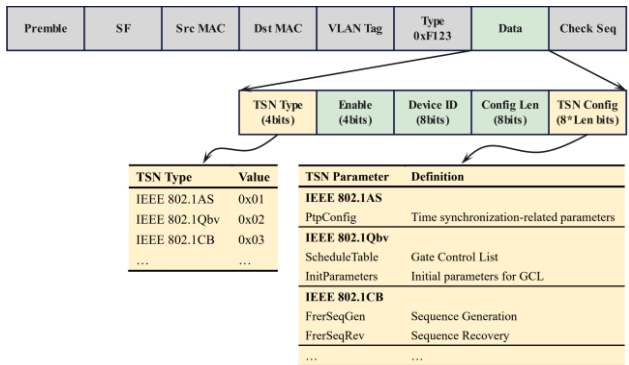

Figure 8. TSN configuration protocol for TSN switch SJA1110.

In addition, built-in topics are predefined in both the CUC and DDS endpoints. Based on these built-in topics, the CUC publishes TSN configuration information to DDS endpoints and notifies them of the exact moment when DDS flows should be released. Similarly, after receiving the TSN configuration information, the DDS endpoints send configuration status to the agent as replies. Once each node returns a message that indicates the configuration was successful, the vehicle enters the run-time phase.

If service updates or upgrades result in changes to DDS flows, the IVNs repeat the network scheduling and network configuring processes to update the TSN configuration of the network devices automatically.

## 6 EXPERIMENTAL RESULTS AND ANALYSIS

### 6.1 Experimental setup

In this section, the performance of DoT is evaluated through physical experiments. The zone-oriented centralized architecture simulated in Figure 9 was set up. In the data plane, zone controllers (ZCU1, ZCU2, and ZCU3) and the central computing unit (CCU) are connected through TSN switches, which form a ring network. The CPUs of the ZCUs and the CCU are NXP S32G399A. The S32G399A contains 8 ARM Cortex-A53 cores and is commonly used in the automotive field as a processor for service-oriented gateways, zone controllers, and vehicle computing platforms. The Linux used in the experiment has a kernel version of 5.10.145 and 3.5GB of memory. The TSN switch is NXP SJA1100, which integrates 1000Base-T1 and 100Base-Tx PHYs. The processing delay of the SJA1110 is less than 10 µs, and the supported TSN protocols include IEEE 802.1AS, IEEE 802.1Qbv, IEEE 802.1Qci, and IEEE 802.1CB.

Table 3. Flow configuration for physical experiments.

| Flow | Size/byte | Period/µs | Priority | Path |
|---|---|---|---|---|
| 1 | 100 | 500 | 7 | ZCU1-> SW2->SW1->CCU |
| 2 | 200 | 500 | 6 | ZCU2->SW3->SW2->SW1->CCU |
| 3 | 150 | 500 | 6 | ZCU3->SW4->SW1 ->CCU |
| 4 | 100 | 300 | 7 | CCU->SW1->SW2->ZCU1 |
| 5 | 100 | 300 | 7 | CCU->SW4->SW3->ZCU2 |
| 6 | 100 | 300 | 7 | CCU->SW1->SW4->ZCU3 |
| 7 | 1000 | 200 | 5 | Cam1->SW3->SW2->SW1->CCU |
| 8 | 1000 | 200 | 5 | Cam2->SW4-> SW1->CCU |
| 9* | 1000 | 50 | BE | ZCU2->SW3->SW2->SW1->CCU |
| 10* | 1000 | 40 | BE | ZCU2->SW3->SW2->SW1->CCU |
| 11* | 1500 | 50 | BE | ZCU2->SW3->SW2->SW1->CCU |
| 12* | 1000 | 80 | BE | ZCU3->SW4->SW3->SW2->SW1->CCU |
| 13* | 1000 | 40 | BE | ZCU3->SW4->SW3->SW2->SW1->CCU |

(Note: Flows marked with an asterisk (*) are interference flows, transmitted only under specific test conditions.)

In the control plane, the CUC, CNC, and agent are the network configuration and management components. The CUC and CNC are deployed on the personal computer. The operating system used is Ubuntu 18.04 running on VMware, with 13.3GB of memory and an Intel i5-13500 CPU (utilizing 8 processor cores). The agent's CPU is a MYC-Y6ULX-V2, running Linux with kernel version 4.9.88 and equipped with 512 MB of memory. In the entire network topology, the network bandwidth for the control plane and the links between the cameras and the TSN switches are 100Mbps, while the bandwidth on other links is 1000Mbps.

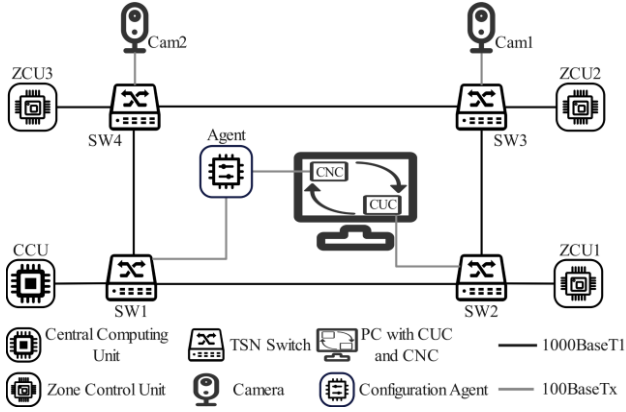

Figure 9. Network topology of physical experiment.

Different DDS applications are deployed on the ZCUs and the CCU to simulate DDS communication in IVNs. DFIA cannot identify video streams generated by cameras, as these cameras do not support DDS. Therefore, video flows are considered static traffic, and their information is pre-configured in the CUC. Table 3 provides details on all flows utilized in the experiment. Additionally, the CUC, CNC, agent, and switches perform the functions described in section 5.

### 6.2 Performance evaluation of FastDDS-lw

Different DDS software was installed on ZCU1 and ZCU2 to evaluate the performance of FastDDS-lw. The performance metrics selected for evaluation include $T_{write}$, $T_{read}$, as well as CPU and memory resource utilization. To ensure the accuracy of the results, no other processes were running on ZCU1 and ZCU2 except the applications for tests, and the network was free of any traffic other than the test flows.

Figure 10 describes the workflow of the experiment. The TestMaster and TestSlave applications were deployed on ZCU1 and ZCU2 respectively. RPC communication was established between TestMaster and TestSlave through two topics, TestReq and TestRep. When TestMaster invokes $write()$ function to publish a test-req message, the system time at that moment is recorded as $t_1$. Upon receiving the test-req message, TestSlave responds with a test-rep message via the TestReply topic. TestMaster then reads the data from the test-rep message using the $read()$ function. The time when the test-rep message is successfully received and deserialized is recorded as $t_4$. The "tcpdump" command was executed on ZCU1 to capture all RTPS messages, with the times of capturing the test-req and test-rep messages recorded as $t_2$ and $t_3$. Consequently, $T_{write}$ and $T_{read}$ are calculated using Equations (6), (7).

$$T_{write} = t_2 - t_1 \qquad (6)$$
$$T_{read} = t_4 - t_3 \qquad (7)$$

The experiments assessed $T_{write}$ and $T_{read}$ of three DDS software under two scenarios: 25% and 75% CPU load(the CPU load was stabilized at these fixed values by executing a specific background process):
1) Fast DDS v2.12.0 with RELIABILITY QoS was configured to "Reliable" (referred to as FastDDS-rel);
2) Fast DDS v2.12.0 with RELIABILITY QoS was configured to "BestEffort" (referred to as FastDDS-be);
3) FastDDS-lw with FRER enabled in SW1 and SW2.

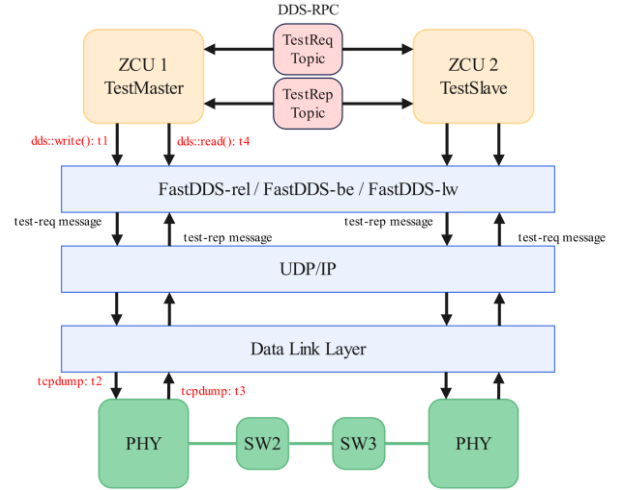

Figure 10. The procedure of performance evaluation experiment.

Each test-req or test-rep message had a size of 1454 bytes, and each experiment was repeated 1000 times. Table 4 presents the average and maximum values for $T_{write}$ and $T_{read}$, CPU usage, and memory usage. Due to the retransmission mechanism, FastDDS-rel exhibited higher processing latency compared to FastDDS-be and FastDDS-lw, with $T_{read}$ reaching the millisecond level when CPU load is 75%. FastDDS-lw achieved reliable communication via FRER while reducing the average values of $T_{write}$ and $T_{read}$ by over 30% compared to FastDDS-rel, and the maximum values by more than 60%. Even when compared to FastDDS-be, which operates on a best-effort basis, FastDDS-lw reduced the average values of $T_{write}$ and $T_{read}$ by up to 11%, and the maximum values by up to 34%. CPU usage for all three DDS software was within 5%, and minimal differences observed. In terms of memory usage, FastDDS-lw consumed approximately 12% less than both FastDDS-be and FastDDS-rel. These results indicate that, with the FRER enabled, FastDDS-lw maintains lower processing latency and smaller jitter compared to both FastDDS-be and FastDDS-rel, while also demonstrating reduced memory consumption.

Table 4. $T_{write}$, $T_{read}$, CPU and memory usage of three DDS software implementations.

| DDS software | | FastDDS-lw | | FastDDS-rel | | | | FastDDS-be | | | |
|---|---|---|---|---|---|---|---|---|---|---|---|
| CPU load (%) | | 25 | 75 | 25 | Δ*(%) | 75 | Δ*(%) | 25 | Δ*(%) | 75 | Δ*(%) |
| $T_{write}$ (μs) | mean | 26.231 | 26.652 | 41.516 | 36.82 | 44.212 | 39.72 | 28.383 | 7.58 | 28.211 | 5.53 |
| | max | 39.216 | 41.808 | 124.833 | 68.59 | 182.737 | 77.12 | 41.024 | 4.41 | 50.184 | 16.69 |
| $T_{read}$ (μs) | mean | 34.469 | 35.021 | 51.61 | 33.21 | 65.398 | 46.45 | 38.744 | 11.03 | 38.874 | 9.91 |
| | max | 41.76 | 58.8 | 242.322 | 82.77 | 2279.403 | 97.42 | 45.512 | 8.24 | 89.632 | 34.40 |
| CPU usage (%) | mean | 1.104 | 1.063 | 1.327 | 16.80 | 1.365 | 22.12 | 1.2 | 8.00 | 1.253 | 15.16 |
| | max | 1.663 | 1.663 | 3.338 | 50.18 | 4.163 | 60.05 | 2.35 | 29.23 | 3.338 | 50.18 |
| Memory usage (KB) | value | 10908 | 11000 | 12536 | 12.99 | 12452 | 12.25 | 12484 | 12.62 | 12484 | 11.89 |

(Note: Δ* represents the percentage reduction in FastDDS-lw results compared to FastDDS-rel and FastDDS-be.)

### 6.3 Performance Evaluation of DFIA

DFIA utilizes a rule-based matching algorithm to identify DDS flows. Provided that all DDS Endpoints adhere to interoperability specifications, DFIA can achieve 100% accuracy. Consequently, this study does not concentrate on the accuracy of DFIA but rather evaluates its processing time. The experiments are organized into two groups:
1) 1 DataWriter and 4 DataReaders(total 4 DDS flows);
2) 1 DataWriter and 8 DataReaders(total 8 DDS flows).

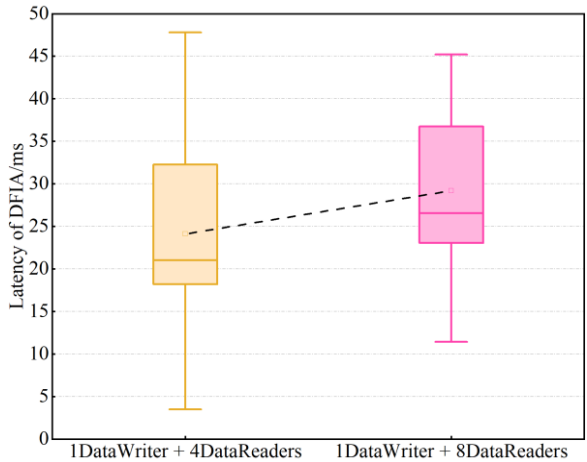

Figure 11. Processing latency of DFIA.

In experiment 1), the DataWriter and DataReaders had the same topic and compatible QoS policies, i.e., the mutual matching rules were satisfied. The Discovery Server, integrated with DFIA, was first activated on the CCU. Subsequently, one DataReader process was initiated on each ZCU and the CCU. Finally, the DataWriter was activated on the CCU. The processing latency, measured from when DFIA identified the DataWriter to when it generated an XML file containing the information for the four DDS flows, was recorded. In experiment 2), two DataReader processes were deployed on each ZCU and the CCU, and the experimental procedure was similar to 1). The results of 100 experiments are illustrated in Figure 11.

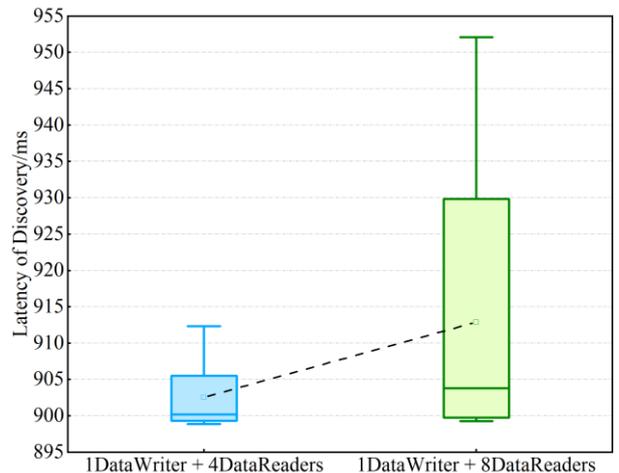

Figure 12. Processing latency of DDS Discovery.

As a control measure, each experiment recorded the time required for DataWriter and DataReaders to discover each other. The results are illustrated in Figure 12. As the number of DataReaders increases, the count of executing the matching algorithm and the number of DDS flows that need to be written into the XML file also increase, leading to a corresponding rise in DFIA's processing latency. DFIA runs on a separate thread, allowing it to execute in parallel with Discovery Server. DFIA's processing latency can be maintained within 50 milliseconds, whereas the discovery process latency typically exceeds 900 milliseconds. Therefore, DFIA can identify DDS flows before the end of the Discovery phase.

### 6.4 Performance Evaluation of DoT

In the run-time phase, best-effort DDS flows of 1000 bytes are randomly injected to verify the communication performance and network configuration function of DoT, which are considered interference traffic. The experiment focuses on the end-to-end latency of flow 1 with a priority of 7. The end-to-end latency of flow 1 is tested under the following five conditions:

1) No interference traffic (as the control group);
2) Interference traffic with a bandwidth consumption of 300Mbps, traditional Ethernet;
3) Interference traffic with a bandwidth consumption of 300Mbps, with TAS enabled;
4) Interference traffic with a bandwidth consumption of 800Mbps, traditional Ethernet;
5) Interference traffic with a bandwidth consumption of 800Mbps, with TAS enabled.

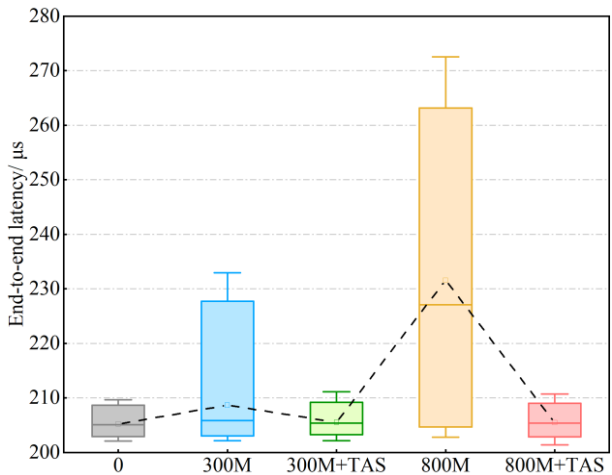

Figure 13. End-to-end latency of DDS Flow 1 in different traffic scenarios.

Each set of experiments was repeated 1000 times. The results are shown in Figure 13. When traditional Ethernet is used, the mean and jitter of the end-to-end latency of DDS flow 1 are significantly affected by the injection of interference traffic. However, when the TAS is enabled, the end-to-end latency of DDS flow 1 remains consistent with that of the control group, regardless of the amount of bandwidth occupied by the interference traffic. The results indicate that, in the DoT architecture, the TAS ensures deterministic and low end-to-end latency for DDS flow.

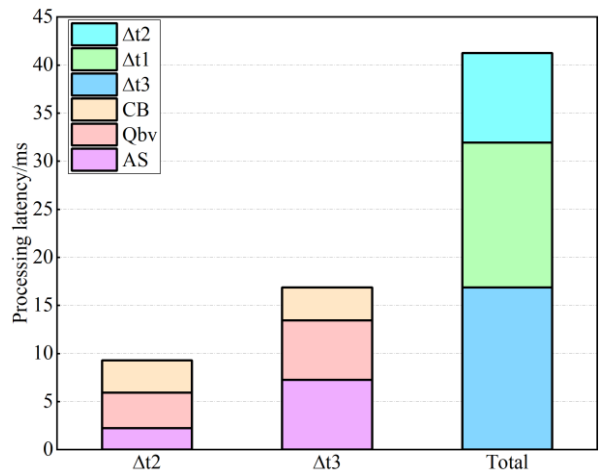

Figure 14. Processing latency at different stages of DoT.

When the bandwidth of interference traffic changes, it is first identified by DFIA, which then triggers the automatic network scheduling and configuring. The processing latency of network scheduling and configuration is also an important metric for evaluating DoT performance. Network configurations for switches and DDS endpoints are usually performed in parallel. For DDS endpoints, the only required configuration is the publication time of DDS flows. The processing latency for configuring endpoints is in the range of hundreds of microseconds, which is significantly shorter than the time required for switch configuration. Therefore, only the processing latency related to switches is recorded:

1) $\Delta_{t1}$: The time required for the CNC to re-execute network scheduling;
2) $\Delta_{t2}$: The time required for the agent to parse the TSN configuration files and send the configuration frames;

3) $\Delta_{t3}$: The time required for 4 switches to update the TSN configuration and return the configuration status.

The experiment is conducted with interference traffic bandwidth increasing from 0M to 800M. The TSN configuration includes AS, Qbv, and CB. The $\Delta_{t2}$ and $\Delta_{t3}$ for each protocol is recorded separately. Each set of experiments is repeated 100 times, and the average processing latency is shown in Figure 14.

The experimental results indicate that the time required for DoT to re-execute the network scheduling and configuring exceeds 40ms. Considering DFIA, the total time required is over 70ms, which means that DoT can automatically complete the network scheduling and configuration before the discovery phase concludes.

**7 Conclusion**

In this paper, a DoT communication architecture was designed to integrate DDS and TSN in IVNs. At first, a lightweight DDS software, FastDDS-lw, was developed based on the actual needs of IVNS. Experimental results show that FastDDS-lw significantly reduces average and maximum processing latency compared to Fast DDS v2.12.0, and achieves a 12% reduction in runtime memory usage. Then, a CDA-based DDS flow identification algorithm was proposed to realize the automatic identification of DDS flows. DFIA can complete the identification of flows before the end of the discovery phase of DDS, without affecting DDS communication. The DoT architecture, utilizing FastDDS-lw and DFIA, aligns with the fully centralized model defined in IEEE 802.1Qcc. Performance evaluations demonstrate that the DoT effectively prevents DDS flows from being blocked even under substantial interfering traffic. Furthermore, DoT offers an automated network configuration method to support flexible service deployment in SOA. In the experimental scenarios, network configuration required only tens of milliseconds. Future work will focus on deploying standard network management and configuration protocols, such as NETCONF, on in-vehicle TSN switches to further streamline the process of network configuration.

**ORCID**

Yi Ren, http://orcid.org/0000-0002-4853-2554

**ACKNOWLEDGEMENTS**

This work was supported by the Shanghai Pudong New Area Science and Technology Development Fund, Industry-University-Research Special Project (Future Vehicle) (No. PKX2022-W01).